\documentclass[conference]{IEEEtran}
\IEEEoverridecommandlockouts
\usepackage{cite}
\usepackage{makecell}
\usepackage[colorlinks=true, allcolors=blue]{hyperref}
\usepackage{amsmath,amssymb,amsfonts}
\usepackage{algorithmic}
\usepackage{booktabs}
\usepackage{graphicx}
\usepackage{textcomp}
\usepackage{xcolor}
\usepackage{float}
\usepackage{multirow}
\usepackage{array}
\usepackage{pifont}
\usepackage{listings}
\usepackage{cite,balance}

\def\BibTeX{{\rm B\kern-.05em{\sc i\kern-.025em b}\kern-.08em
    T\kern-.1667em\lower.7ex\hbox{E}\kern-.125emX}}
\begin{document}

\title{Overview of ESDD2: Environment-Aware\\ Speech and Sound Deepfake Detection Challenge}

\author{
Xueping Zhang$^{1}$, Han Yin$^{2}$, Yang Xiao$^{3}$, Lin Zhang$^{4}$, Ting Dang$^{3}$, \\ Rohan Kumar Das$^{5}$, Ming Li$^{6}$ \\
$^{1}$Duke Kunshan University 
$^{2}$Korea Advanced Institute of Science and Technology, \\
$^{3}$The University of Melbourne, 
$^{4}$Johns Hopkins University,
$^{5}$Fortemedia Singapore\\
$^{6}$The Chinese University of Hong Kong, Shenzhen \\
\url{https://sites.google.com/view/esdd-challenge/esdd-challenges/esdd-2/description}
}

\maketitle

\begin{abstract}
Audio recorded in real‑world environments often contains a mixture of components:  foreground speech and background environmental sounds. With rapid advances in text‑to‑speech, voice conversion, and other generation models, either component can now be modified independently. 
Such component-level manipulations are harder to detect, as the remaining unaltered component can mislead the systems designed for whole deepfake audio, and they often sound more natural to human listeners.
To address this gap, we are launching the second Environment‑Aware Speech and Sound Deepfake Detection Challenge (ESDD2), which focuses on defending against more challenging and realistic environment‑aware, component‑level deepfake manipulations. ESDD2 is built on our recently released CompSpoofV2 dataset and uses a separation‑enhanced joint learning framework as the baseline.
This paper presents an overview of the ESDD2 Challenge: the task description, the new database, baseline, metric, and a summary of the results.
\end{abstract}


\section{Introduction}
In real-world recording scenarios, audio often contains a mixture of foreground speech and background sounds. We define such audio as comprising two components: (i) speech, i.e., linguistically meaningful speech produced by the primary foreground speaker, and (ii) environmental sound, i.e., any non‑speech background or non‑target speech. With rapid advances in text‑to‑speech, voice conversion, and other generation models, either component can now be modified independently. For example, replacing the background while leaving the foreground human speech unchanged, or modifying the speech content while preserving the background.
Such component-level manipulations are harder to detect, as the remaining unaltered component can mislead the systems designed for whole deepfake audio, and they often sound more natural to human listeners.

To address this gap, we introduced a component-level spoofing audio dataset CompSpoof~\cite{zhang2025compspoof} in 2025. Then, we extend the dataset to CompSpoofV2, which contains more than 250,000 audio clips (around 283 hours) formed by mixing bona fide (genuine or real) and spoofed (synthetic or manipulated) audio, including speech and environmental sounds, from multiple sources. We also proposed a separation-enhanced joint learning framework to address this gap.

To further promote the development of environment-aware speech and
sound deepfake detection, we extend our first ESDD challenge\cite{ESDD1,yin2025esdd,yin2026first} and launch the second Environment-Aware Speech and Sound Deepfake Detection Challenge (ESDD2) at ICME 2026. While our first ESDD challenge focuses on the detection of forged environmental sounds, ESDD2 advances this direction by addressing environment-aware component-level spoofing, where either or both speech and environmental sounds may be manipulated or synthesized, posing a more challenging and realistic detection scenario.

The main contributions are summarized as follows:
\begin{itemize}
 \item We present a comprehensive description of the ESDD2 challenge, including the dataset, task description, baseline systems and evaluation criteria.
 \item We report and analyze the leaderboard results, providing a systematic comparison of participating systems.
 \item We summarize common effective design choices observed in top-performing systems, offering insights for future research on environment-aware speech and sound deepfake detection.
\end{itemize}

\begin{table*}[h]
\caption{Five Categories Considered in CompSpoofV2 dataset.}
{
\centering{
\begin{tabular}{ c c c c c l}
\toprule
\textbf{ID} & \textbf{Mixed} & \textbf{Speech} & \textbf{Environment} & \textbf{Class Label} & \textbf{Description} \\
\midrule
0 & \ding{55} & Bona fide & Bona fide &original & Original audio without any manipulation or mixing\\
1 & \ding{51} & Bona fide & Bona fide &bonafide\text{\_}bonafide & Bona fide speech mixed with bona fide environmental sound from other audio\\
2 & \ding{51} & Spoofed   & Bona fide &spoof\text{\_}bonafide & Spoofed speech mixed with bona fide environmental sound \\
3 & \ding{51} & Bona fide & Spoofed   &bonafide\text{\_}spoof & Bona fide speech mixed with spoofed environmental sound \\
4 & \ding{51} & Spoofed   & Spoofed   &spoof\text{\_}spoof & Spoofed speech mixed with spoofed environmental sound \\
\bottomrule
\end{tabular}
}
}
\label{tab:CompSpoof_classes}
\end{table*}

\begin{table*}[h]
\centering
\caption{Audio sources for the training and validation sets.}
\label{tab:audio_sources_train}
\setlength{\tabcolsep}{2pt} 
{\fontsize{6.9pt}{10pt}\selectfont
\begin{tabular}{@{}l|@{}ccc@{}}
\toprule
\textbf{ID} & \textbf{Original Source} & \textbf{Speech Source} & \textbf{Environmental Sound Source} \\
\midrule
0 & AudioCaps\cite{AudioCaps}, VGGSound\cite{VGGSound} & -- & -- \\
1 & -- & CommonVoice\cite{CommonVoice}, LibriTTS\cite{LibriTTS}, english-conversation-corpus\cite{ecc} & AudioCaps, TAUUAS\cite{TAUUAS}, TUTSED\cite{TUTSED2016, TUTSEDdev2017, TUTSEDeval2017}, UrbanSound\cite{UrbanSound}, VGGSound \\
2 & -- & CommonVoice, LibriTTS & EnvSDD\cite{EnvSDD}, VcapAV\cite{VcapAV} \\
3 & -- & ASV5\cite{ASV5}, MLAAD\cite{MLAAD} & AudioCaps, TAUUAS, TUTSED, UrbanSound, VGGSound \\
4 & -- & ASV5, MLAAD & EnvSDD, VcapAV \\
\bottomrule
\end{tabular}
}
\end{table*}

\begin{table*}[h]
\centering
\caption{Audio sources for the evaluation and test sets.}
\label{tab:audio_sources_eval}
\begin{tabular}{l|ccc}
\toprule
\textbf{ID} & \textbf{Original Source} & \textbf{Speech Source} & \textbf{Environmental Sound Source} \\
\midrule
0 & AudioCaps, VGGSound & -- & -- \\
1 & -- & CommonVoice, LibriTTS, english-conversation-corpus & AudioCaps, TAUUAS, TUTSED, UrbanSound, VGGSound \\
2 & -- & CommonVoice, LibriTTS & EnvSDD, VcapAV, \textbf{New Generated by MeanAudio} \\
3 & -- & ASV5, MLAAD, \textbf{New Generated by SoulX-Podcast} & AudioCaps, TAUUAS, TUTSED, UrbanSound, VGGSound \\
4 & -- & ASV5, MLAAD, \textbf{New Generated by SoulX-Podcast} & EnvSDD, VcapAV, \textbf{New Generated by MeanAudio} \\
\bottomrule
\end{tabular}
\end{table*}

\section{ESDD2 Challenge}

\subsection{Task Description}
\begin{figure}[!h]
\centerline{\includegraphics[width=0.5\textwidth]{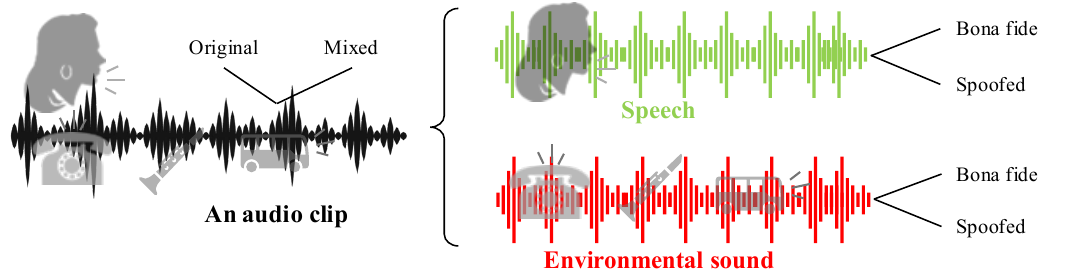}}
\caption{ESDD2 Task illustration. An audio clip is first classified as mixed or original; for mixed audio, speech and environmental sound components are separately evaluated for genuineness, resulting in five target classes.}
\label{main1}
\end{figure}
In ESDD2 and the discussed CompSpoofV2, we consider the scenario where each audio belongs to one of five categories, covering all combinations of bona fide and spoofed speech and environmental sound as shown in Table \ref{tab:CompSpoof_classes}. The task of this challenge is to classify a given audio clip into one of the five defined classes, as shown in Figure \ref{main1}.

The training and validation sets share the same data sources and class distribution.
Similarly, the evaluation and test sets share the same data sources and class distribution.
Notably, the evaluation and test sets include newly generated audio samples that are unseen in the training and validation sets.

\subsection{CompSpoofV2 Dataset}

\begin{table*}[h]
\caption{Performance of the Top-12 Systems (F1-score / Original EER / Speech EER / Environmental EER) ``Ensem'' denotes the number of models in an ensemble (``1'' indicates a single model). }
\begin{tabular}{@{}c@{}|l|@{}l@{}|@{}c@{}|l@{}|l@{}|l|l@{}}
\toprule
\multirow{2}{*}{\textbf{\begin{tabular}[c]{@{}l@{}}Team \\ No.\end{tabular}}} & \multirow{2}{*}{\textbf{\begin{tabular}[c]{@{}l@{}}Team \\ Affiliation\end{tabular}}} & \multirow{2}{*}{\textbf{System}} & \multirow{2}{*}{\textbf{\begin{tabular}[c]{@{}l@{}}Ensem\end{tabular}}} & \multirow{2}{*}{\textbf{Param.}} & \multirow{2}{*}{\textbf{Data Aug.}} & \multirow{2}{*}{\textbf{Eval set}}         & \multirow{2}{*}{\textbf{Test set}}         \\
                                                                              &                                                                                       &                                  &                                                                                &                                  &                                     &                                            &                                            \\ \hline
1                                                                             & AHU                                                                                   & E2E-EA-SSDD                      & 7                                                                              & 6.56B                            & (1)(2)                              & -                                          & 0.8775 /      -       /      -      /    - \\
2                                                                             & CUC                                                                                   & EnvTriCascade                    & 2                                                                              & 540.81M                          & (1)                                 & -                                          & 0.8266 /      -       /      -      /    - \\
3                                                                             & SETW                                                                                  & FrozenSSL-Ens4                   & 4                                                                              & 1908M                            & (1)(3)(4)(5)                        & 0.8124 /      -       /      -      /    - & 0.8200 /      -       /      -      /    - \\
4                                                                             & HKUST(GZ)                                                                             & GLADSE                           & 2                                                                              & 674.57M                          & (1)                                 & 0.7120 / 0.0102 / 0.1201 / 0.0999          & 0.8077 / 0.0133 / 0.0896 / 0.0926          \\
5                                                                             & SIT                                                                                   & CompEnsFusion                    & 8                                                                              & 4.6B                             & (1)                                 & 0.7715 / 0.0124 / 0.2177 / 0.1341          & 0.7828 / 0.0109 / 0.2164 / 0.1263          \\
6                                                                             & WHU                                                                                   & LaMSep-DF                        & 1                                                                              & 356.85M                          & (1)                                 & 0.7045 / 0.0227 / 0.2889 / 0.2218          & 0.7262 / 0.0150 / 0.1171 / 0.0869          \\
7                                                                             & JAIST\_1                                                                                 & MIF                              & 4                                                                              & 1034.15M                         & -                                   & 0.6995 / 0.0560 / 0.1852 / 0.1436          & 0.7187 / 0.0345 / 0.1562 / 0.1100          \\
8                                                                             & SCISTOR                                                                               & HeteroSSL-Fus                    & 1                                                                              & 440M                             & (1)(3)(4)(5)                        & -                                          & 0.7137 / 0.0314 / 0.3010 / 0.1594          \\
9                                                                             & JAIST\_2                                                                               & EAT-XLSR-MTL                     & 2                                                                              & 523.09 M                         & (6)                                 & 0.6274 / 0.0977 / 0.2349 / 0.1470          & 0.7124 / 0.0116 / 0.2604 / 0.1703          \\

10                                                                             & XHU                                                                                  & CAFM-MTL                         & 2                                                                              & -                         & -                                 & -         & 0.7056 /  0.0172 / 0.1053 / 0.1228          \\

11                                                                            & JAIST\_3                                                                                 & XLSR+BEATs+MH\cite{t11}                    & 1                                                                              & 398.140M                         & (3)(5)(7)(8)                        & 0.7011 / 0.0299 / 0.3140 / 0.1654          & 0.7019 / 0.0259 / 0.3298 / 0.1883          \\
12                                                                            & NBU                                                                                   & Feature Decomp.                  & 1                                                                              & 654.00M                          & -                                   & 0.5962 / 0.0179 / 0.4489 / 0.4568          & 0.6977 / 0.0172 / 0.2307 / 0.2853          \\
13                                                                            & IITJ                                                                                  & CompMulTask                      & 1                                                                              & 957M                             & (9)(10)(11)(12)                     & 0.6828 / 0.0632 / 0.1599 / 0.1155          & 0.6840 / 0.0644 / 0.1587 / 0.1126          \\ \midrule
-                                                                             & Baseline                                                                              & Separation + AASIST              & 1                                                                              & 957.85M                          & -                                   & 0.6224 / 0.0174 / 0.1993 / 0.4336          & 0.6327 / 0.0173 / 0.1978 / 0.4279          \\ \midrule
\multicolumn{8}{l}{\textbf{Data Aug. Methods:}}                                                                                                                                                                                                                                                                    \\
\multicolumn{8}{l}{\makecell[l]{(1)Rawboost; (2)Loudness Aug.; (3)Codec Aug.; (4)Volume Perturbation; (5)Additive Noise; (6)SpecAugment; (7)Mixup; (8)Temporal aug.; \\ (9)Random Cropping; (10)Zero-padding; (11)Mini-batch Shuffling; (12)Class-balanced} }                                                                                                                                                                                                                                                                                                                                                                    \\ \bottomrule
\end{tabular}
\label{res_final}
\end{table*}
The \textbf{CompSpoofV2}\footnote{\url{https://xuepingzhang.github.io/CompSpoof-V2-Dataset/}}  is a dataset designed for component-level anti-spoofing detection research, where either the speech or the environmental sound component (or both) may be spoofed.
CompSpoofV2 contains over 250,000 audio samples, with a total duration of approximately 283 hours. Each audio sample has a fixed length of 4 seconds and is provided at multiple sampling rates, enabling a more faithful simulation of real-world acoustic and system-level variations.
Building upon the CompSpoof dataset\cite{zhang2025compspoof}, CompSpoofV2 significantly expands the diversity of attack sources, environmental sounds, and mixing strategies. In addition, newly generated audio samples are distributed across the test set and are specifically designed to serve as detection data under unseen conditions. The audio sources for each category are summarized in Table~\ref{tab:audio_sources_train} and Table~\ref{tab:audio_sources_eval}.

\subsection{Baseline}

\begin{figure}[!h]
\centerline{\includegraphics[width=0.48\textwidth]{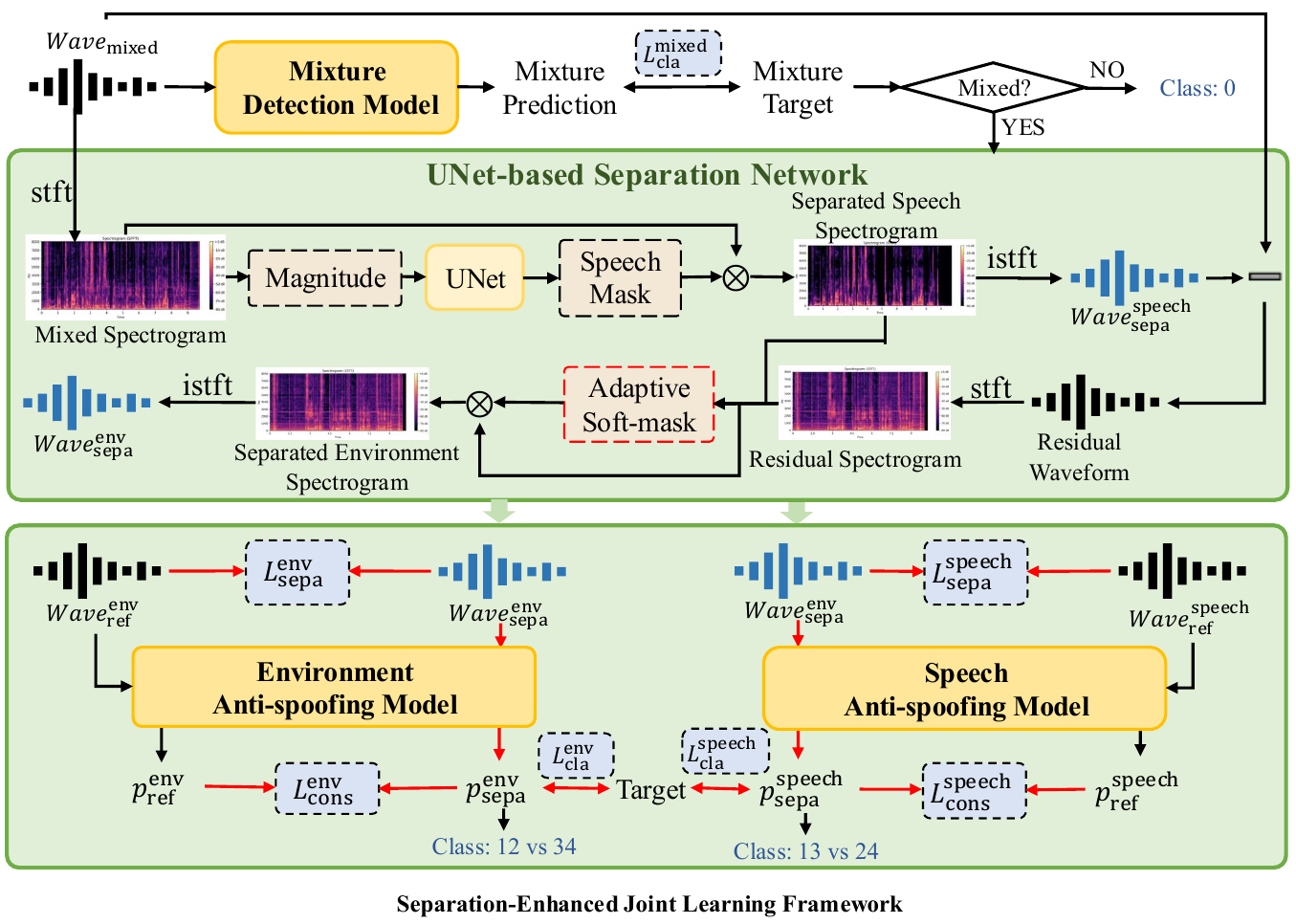}}
\caption{Overview of the proposed separation-enhanced joint learning framework. `\textcolor{red}{$\rightarrow$} ' illustrates the joint learning data flow between the separation and anti-spoofing models.}
\label{main2}
\end{figure}

We utilize a \textbf{separation-enhanced joint learning framework}~\cite{zhang2025compspoof} as the baseline for this task. As shown in Fig.~\ref{main2}, the framework consists of a mixture detection model, a UNet-based separation network, and two component-wise anti-spoofing models for speech and environmental sound. 
The framework first detects potentially spoofed mixtures. If the input audio is detected as containing spoofed mixtures, it will be separated into speech and environmental components. Subsequently, these components are processed by speech- and environment-specific anti-spoofing models. 
Their outputs are fused and mapped to five-class predictions. The separation network operates in the STFT domain, where the speech component is estimated via a learned mask and reconstructed using inverse STFT. The environmental sound is derived from the residual signal with an adaptive soft-masking strategy to suppress speech leakage. The separation network is trained using Cross-Entropy loss on both components.

To ensure spoof-relevant information is preserved, we jointly train the separation and anti-spoofing models. The separated speech and environmental signals are fed into their respective detectors with classification supervision, while a consistency constraint aligns their predictions with those from the reference signals. During inference, the mixture is first classified, then separated into speech and environmental components, which are evaluated independently. Their outputs are combined to produce the final multi-class prediction.

\subsection{Evaluation Criteria}

System performance will be evaluated using the \textbf{Overall Macro-F1 score} across all five target classes, as shown in ~\eqref{eq1}. The \textbf{Macro-F1 score} is computed as the arithmetic mean of the per-class F1 scores. This ensures all classes contribute equally to the final evaluation, regardless of sample imbalance.
\begin{equation}
\label{eq1}
\text{Macro-F1} = \frac{1}{5} \sum_{i=1}^{5} \text{F1}_i, 
\end{equation}
where, the $\text{F1}_i$ score is calculated as the harmonic mean of precision $P_i$ and recall $R_i$ for class $i$:
\begin{equation}
\label{eq2}
\text{F1}_i = \frac{2 \cdot P_i \cdot R_i}{P_i + R_i}, \quad
P_i = \frac{\text{TP}_i}{\text{TP}_i + \text{FP}_i}, \quad
R_i = \frac{\text{TP}_i}{\text{TP}_i + \text{FN}_i}.
\end{equation}
$\text{TP}_i$, $\text{FP}_i$, and $\text{FN}_i$ denote the number of true positives, false positives, and false negatives for class $i$, respectively.
A higher Macro-F1 score indicates better overall system performance.

In addition, we introduce three auxiliary metrics based on equal error rate (EER): $\mathrm{EER}_{\text{original}}$, $\mathrm{EER}_{\text{speech}}$ and $\mathrm{EER}_{\text{env}}$.
$\mathrm{EER}_{\text{original}}$ is the EER for distinguishes the \texttt{Original} ($i=0$) class from all other four classes $i=\{1,2,3,4\}$. Similarly, $\mathrm{EER}_{\text{speech}}$ is the EER for detecting spoofed speech component, while $\mathrm{EER}_{\text{env}}$ is the EER for detecting spoofed environmental component. The three EER metrics are included for diagnostic and analysis purposes only and are \textbf{not} considered for the leaderboard ranking.

\section{Results and Discussions}

We received 94 registrations from 16 countries. The initial leaderboard contained additional submissions. After verifying the required metadata and challenge submission requirements, 13 teams were retained in the final leaderboard for the analysis. The challenge consists of two phases: a \textit{Preparation Phase} using the evaluation set, and a \textit{Final Ranking Phase} using the test set. The final rankings are determined based on the F1-score in the Final Ranking Phase as present in Table \ref{res_final}.

\subsection{Overall System Trends}
The results reveal that the best systems do not rely on one backbone. They combine different models and split the task into clear sub-problems. Top teams often use larger models, but a bigger size does not always mean better results. Team No.1 ranks first with a 6.56B ensemble. Team No.2 ranks second with a much smaller 540.81M cascaded system. This means model complementarity and good decision design matter more than parameter count alone. Efficiency is also important. Many teams improved the baseline while reduced the number of parameters. They saved computation and memory while keeping better performance. This shows better architecture design can outperform simple model scaling.

\subsection{Cross-Domain SSL Backbones}
Another clear trend is the use of cross-domain self-supervised learning (SSL) backbones. Besides common speech SSL encoders like XLS-R\cite{xlsr}, top systems use newer models such as EAT\cite{eat}, SSLAM\cite{sslam}, Dasheng\cite{dasheng}, and DF-Arena\cite{DF-Arena}. These models provide different strengths. Speech-focused encoders capture phonetic and prosodic artifacts. Audio/event-focused encoders are better at environmental mismatch and mixture traces. Team No.2 and No.3 use this idea directly by combining EAT/SSLAM-type branches with XLS-R branches. Team No.5 combines DF-Arena with XLSR-Mamba\cite{XLSR-Mamba}, SLS\cite{sls}, and TCM-ADD\cite{TCM-ADD}, then applies component-wise fusion and post-hoc calibration.

\subsection{Ensemble and Fusion Strategies}
We also observe that ensemble quality matters more than ensemble size. Team No.5 uses 8 checkpoints, but team No.3 with 4 models and team No.2 with 2  models achieve stronger or similar performance. This suggests that a small set of diverse experts with clear roles is better than a large homogeneous ensemble. 

\subsection{Data Augmentation}

Data augmentation is still a key factor. RawBoost\cite{rawboost} is the most common method in high-ranked systems. It is often combined with codec simulation, additive noise, and volume perturbation. These methods are especially useful for robust environmental spoofing detection.

\subsection{Architectural Diversity}
Importantly, gains do not come only from augmentation or large ensembles. From the system architecture perspective, the top teams do not follow one single architectural style. Instead, they cover multiple strong routes: direct end-to-end 5-class modeling (e.g., team No.1), cascaded coarse-to-fine pipelines (e.g., team No.1 and team No.9), multi-backbone direct fusion (e.g., team No.9 and team No.7), component-level multi-model ensembling with post-hoc calibration (e.g., team No.9), dual-branch speech–environment interaction modeling (e.g., team No.4 and team No.10), separation-driven designs (e.g., team No.6, team No.8 and team No.12), and feature-disentanglement-based designs (e.g., team No.11). Although these systems differ in pipeline form, they share a practical direction: they improve how speech and environmental cues are modeled and then combine those cues with stronger fusion or decision strategies, rather than relying on a single fixed backbone or a single fixed processing stage.

Overall, top submissions follow a common recipe. They use modular designs and combine multiple backbones, especially audio SSL models. They split the task into simpler sub-problems with cascaded, multi-branch, or separation-based pipelines. They do not rely on large models or big ensembles. Instead, they use a small number of diverse models with clear roles, and apply effective fusion and calibration. They also improve efficiency by refining baseline structures to reduce computation while keeping strong performance. Together with solid data augmentation, these designs show diversity, structure, and efficiency matter more than simple model scaling.

\section{Conclusions and Future Scope}
In this paper, we summarized the ICME ESDD2 Challenge, including the dataset, task description, baseline system, evaluation criteria, and challenge results analysis. The strong performance achieved by the participating teams highlights the progress made in current deepfake detection methods. At the same time, the results reveal remaining challenges in generalization and robustness across diverse sound types and generation settings. In the future, we plan to further expand the dataset with more realistic and diverse deepfake scenarios and encourage the development of more robust and explainable detection approaches.

\section{Acknowledgment}
This challenge is sponsored by OfSpectrum, Inc. \footnote{\url{https://ofspectrum.com/}}. OfSpectrum, Inc. provided a 1,000 USD prize for the first-place winner. The company develops imperceptible watermarking technology for content provenance, copyright protection, and trustworthy AI voice applications.
 
\balance
\bibliographystyle{IEEEbib}
\bibliography{icme2026references}

\end{document}